\documentclass[twocolumn,nofootinbib,superscriptaddress,floatfix,usenames,dvipsnames]{revtex4-2}
\pdfoutput=1

\usepackage[totalwidth=480pt, totalheight=680pt]{geometry}
\usepackage{amsmath, array}
\usepackage{amsfonts}
\usepackage[colorlinks=true, linktoc=page, citecolor=blue, urlcolor=blue, bookmarksopen=true]{hyperref}
\usepackage[retainorgcmds]{IEEEtrantools}
\usepackage{scalerel}

\begin{document}

\title{\color{Blue}\textbf{Cascade of instabilities in the classical limit of the BMN matrix model}}

\author{\textbf{Minos Axenides}}
\email{axenides@inp.demokritos.gr}
\affiliation{Institute of Nuclear and Particle Physics, National Centre for Scientific Research, "Demokritos", 153 10, Agia Paraskevi, Greece}
\author{\textbf{Emmanuel Floratos}}
\email{mflorato@phys.uoa.gr}
\affiliation{Institute of Nuclear and Particle Physics, National Centre for Scientific Research, "Demokritos", 153 10, Agia Paraskevi, Greece}
\affiliation{Department of Nuclear and Particle Physics, National and Kapodistrian University of Athens, 157 84, Athens, Greece}
\author{\textbf{Dimitrios Katsinis}}
\email{dkatsinis@phys.uoa.gr}
\affiliation{Institute of Nuclear and Particle Physics, National Centre for Scientific Research, "Demokritos", 153 10, Agia Paraskevi, Greece}
%\affiliation{Institute of Nuclear and Particle Physics\\ National Centre for Scientific Research, "Demokritos"\\ 153 10, Agia Paraskevi, Greece}
%\affiliation{Department of Nuclear and Particle Physics,\\ National and Kapodistrian University of Athens\\ 157 84, Athens, Greece}
\author{\textbf{Georgios Linardopoulos}}
\email{glinard@inp.demokritos.gr}
\affiliation{Institute of Nuclear and Particle Physics, National Centre for Scientific Research, "Demokritos", 153 10, Agia Paraskevi, Greece}
\begin{abstract}
\vspace{.2cm}\normalsize{\noindent We study the leading (LO) and the next-to-leading order (NLO) stability of multipole perturbations for a static dielectric M2-brane with spherical topology in the 11-dimensional maximally supersymmetric plane-wave background. We observe a cascade of instabilities that originates from the dipole ($j=1$) and quadrupole ($j=2$) sectors (the only unstable sectors of the LO) and propagates towards all the multipoles of the NLO.}
\end{abstract}

\maketitle
\section[Introduction]{Introduction}
\noindent According to our current understanding of the black hole (BH) information paradox, the chaotic dynamics of the microscopic degrees of freedom that are present on BH horizons is an indispensable aspect of its resolution \cite{SekinoSusskind08}. In the present paper, we make an analogy with the BH membrane paradigm \cite{Damour78, ThornePriceMacdonald86} and attempt to describe the dynamics of BH horizons by means of relativistic membranes inside plane-wave spacetimes. In the context of M-theory, these membranes arise from the BMN matrix model in the limit of large matrix dimensions \cite{Hoppe82, DasguptaJabbariRaamsdonk02}. Our goal is to study the emergence of classical chaos at both the leading (LO) and the next-to-leading order (NLO) of perturbation theory for various spherical membranes, by using methods from chaos theory and dynamical systems. \\
\indent Our previous findings \cite{AxenidesFloratosLinardopoulos17a, AxenidesFloratosLinardopoulos17b, AxenidesFloratosKatsinisLinardopoulos20} imply that the various modes of multipole perturbations are coupled to each other. This effect is responsible for the propagation (or the cascade) of dipole and quadrupole instabilities from LO perturbation theory to higher-order multipoles (which are stable to LO). There is a striking similarity between this cascade of instabilities and the avalanche phenomenon (i.e.\ the breaking of large vortices into smaller ones) that characterizes turbulent flows in hydrodynamics (see e.g.\ \cite{Frisch95}). \\
\indent Our paper is organized as follows. In section \ref{Section:Setup} we review the $SO(3)$ symmetric ansatz which is our main object of study. In section \ref{Section:LeadingOrderPT} we set up and carry out the LO perturbative analysis for this system, filling out and supplying many details that were omitted in our previous works. In section \ref{Section:HigherOrderPT} we proceed to the study of higher-order perturbations that lead to the instability cascade phenomenon. An illustrative example of instability cascades can be found in section \ref{Section:Example}. A brief discussion of our findings as well as the conclusions have been included in section \ref{Section:ConclusionsDiscussion}.
\section[Setup]{Setup \label{Section:Setup}}
\noindent One of the most remarkable properties of plane-wave spacetimes is that they can be obtained from any given metric via the Penrose limiting procedure \cite{Penrose76}, which consists in blowing up the spacetime around null geodesics (effectively "zooming in" to them). As it turns out, plane-waves preserve the supersymmetries of the original background, so that the maximally supersymmetric spacetimes AdS$_{4/7}\times\text{S}^{7/4}$ of 11-dimensional supergravity give rise to the maximally supersymmetric plane-wave solution \cite{OFarrillPapadopoulos02b}:
\begin{IEEEeqnarray}{ll}
ds^2 = &\hspace{-.3cm} -2 dx^{+} dx^{-} + \sum_{i=1}^3 dx_i dx_i + \sum_{k=1}^6 dy_k dy_k - \nonumber \\
&\hspace{-.3cm} - \left[\frac{\mu^2}{9}\sum_{i=1}^3 x_i x_i + \frac{\mu^2}{36}\sum_{k=1}^6 y_k y_k\right] dx^+ dx^+ \qquad \label{MaximallySupersymmetricMetric} \\
F_{123+} = &\mu, \label{MaximallySupersymmetricMetricFieldStrength}
\end{IEEEeqnarray}
where $F_{\mu\nu\rho\sigma}$ is the field strength tensor of 11-dimensional supergravity. \\
\indent The Hamiltonian of a bosonic relativistic membrane in the 11-dimensional maximally supersymmetric plane-wave background \eqref{MaximallySupersymmetricMetric}--\eqref{MaximallySupersymmetricMetricFieldStrength} reads in the so-called light-cone gauge ($x^+ \equiv \tau$):
\begin{IEEEeqnarray}{l}
H = \frac{T}{2}\int_{\Sigma} d^2\sigma\bigg[p_x^2 + p_y^2 + \frac{1}{2}\left\{x_i,x_k\right\}^2 + \frac{1}{2}\left\{y_i,y_k\right\}^2 + \nonumber \\
+ \left\{x_i,y_k\right\}^2 + \frac{\mu^2 x^2}{9} + \frac{\mu^2 y^2}{36} - \frac{\mu}{3}\,\epsilon_{ikl}\left\{x_i,x_k\right\}x_l\bigg], \quad \label{ppWaveHamiltonian}
\end{IEEEeqnarray}
where $T$ is the membrane tension and the indices of the coordinates $x$ and $y$ run from 1 to 3 and 1 to 6 respectively. The corresponding dynamics is described by the Gauss law constraint,
\begin{IEEEeqnarray}{c}
\left\{\dot{x}_i, x_i\right\} + \left\{\dot{y}_k, y_k\right\} = 0, \label{GaussLaw1}
\end{IEEEeqnarray}
and the equations of motion,
\begin{IEEEeqnarray}{ll}
\ddot{x}_i = &\left\{\left\{x_i,x_k\right\},x_k\right\} + \left\{\left\{x_i,y_k\right\},y_k\right\} - \frac{\mu^2}{9}\,x_i + \qquad \nonumber \\
& + \frac{\mu}{2}\epsilon_{ikl}\left\{x_k,x_l\right\}, \label{xEquation1} \\
\ddot{y}_i = &\left\{\left\{y_i,y_k\right\},y_k\right\} + \left\{\left\{y_i,x_k\right\},x_k\right\} - \frac{\mu^2}{36}\,y_i. \label{yEquation}
\end{IEEEeqnarray}
\indent It can be shown that the regularized light-cone supermembrane \eqref{ppWaveHamiltonian} is described by the Berenstein-Maldacena-Nastase (BMN) matrix model \cite{BMN02}. Equivalently, the Hamiltonian \eqref{ppWaveHamiltonian} describes the continuum limit of the BMN matrix model \cite{DasguptaJabbariRaamsdonk02}. In the absence of background flux ($\mu = 0$), the plane-wave \eqref{MaximallySupersymmetricMetric}--\eqref{MaximallySupersymmetricMetricFieldStrength} becomes flat 11-dimensional spacetime and the BMN matrix model reduces to the matrix model of Banks, Fischler, Shenker and Susskind (BFSS) \cite{BFSS97}. \\
\indent The presence of a nonzero 4-form field strength $F_{\mu\nu\rho\sigma}$ in the plane-wave background \eqref{MaximallySupersymmetricMetricFieldStrength} induces repulsive flux terms in the membrane effective potential. The potential also has attractive quartic and quadratic terms. These are respectively induced by the self-interaction terms of the M2-brane action and the mass terms of the plane-wave metric \eqref{MaximallySupersymmetricMetric}. The competition of attractive and repulsive forces gives rise to stable dielectric membrane solutions that are named after Myers who studied a similar effect \cite{Myers99b}. \\
\indent A static dielectric Myers solution of spherical topology that lives exclusively in the $SO(3)$ symmetric subspace of the background \eqref{MaximallySupersymmetricMetric}--\eqref{MaximallySupersymmetricMetricFieldStrength} can be obtained from the following ansatz:\footnote{See \cite{AxenidesFloratosKatsinisLinardopoulos20} for a generalized form of this solution. A similar ansatz was studied in \cite{HarmarkSavvidy00}, although the prototype was probably introduced a long time ago \cite{CollinsTucker76}.}
\begin{IEEEeqnarray}{c}
x_i = \mu u_i e_i, \ i = 1,2,3, \qquad y_k = 0, \ k = 1,\ldots 6, \qquad \label{StaticAnsatz}
\end{IEEEeqnarray}
where the three coordinates of the unit sphere,
\begin{IEEEeqnarray}{c}
(e_1, e_2, e_3) = (\cos\phi \sin\theta, \sin\phi \sin\theta, \cos\theta), \qquad \label{Epsilon1}
\end{IEEEeqnarray}
satisfy the $\mathfrak{so}\left(3\right)$ Poisson bracket algebra and are orthonormal:
\begin{IEEEeqnarray}{c}
\{e_a, e_b\} = \epsilon_{abc} \, e_c, \qquad \int e_a \, e_b\,d^2\sigma = \frac{4\pi}{3} \, \delta_{ab}. \qquad \label{Epsilon2}
\end{IEEEeqnarray}
The membrane effective potential that arises when the ansatz \eqref{StaticAnsatz} is inserted into the Hamiltonian \eqref{ppWaveHamiltonian} is the generalized 3d H\'{e}non-Heiles potential \cite{EfstathiouSadovskii04}:
\begin{IEEEeqnarray}{ll}
V_{\text{eff}} = \frac{2\pi T \mu^4}{3}\bigg[&u_1^2 u_2^2 + u_2^2 u_3^2 + u_1^2 u_3^2 + \nonumber \\
& + \frac{1}{9}\left(u_1^2 + u_2^2 + u_3^2\right) - 2 u_1 u_2 u_3\bigg]. \qquad \label{StaticPotential}
\end{IEEEeqnarray}
\eqref{StaticPotential} has exactly nine critical points \cite{AxenidesFloratosLinardopoulos17b}. These are
\begin{IEEEeqnarray}{c}
\textbf{u}_0 = 0, \ \textbf{u}_{1/6} = \frac{1}{6}\,\left(1, 1, 1\right), \ \textbf{u}_{1/3} = \frac{1}{3}\,\left(1, 1, 1\right), \qquad \label{So3Extrema1}
\end{IEEEeqnarray}
as well as six more extrema that are obtained from $\textbf{u}_{1/6}$ and $\textbf{u}_{1/3}$ by flipping the signs of two of their components. The discrete symmetry group of the nine extremal points of \eqref{StaticPotential} is the tetrahedral group $T_d$. This discrete symmetry is shared by the equations of motion \eqref{xEquation1}--\eqref{GaussLaw1}. To see this, apply separate reflections to the coordinates $x_i \rightarrow \epsilon_i x_i$ ($\epsilon_i = \pm 1$) and note that the flux term implies the constraint $\epsilon_1\epsilon_2\epsilon_3 = 1$. \\
\indent In sum, the potential \eqref{StaticPotential} has degenerate minima at $\textbf{u}_0$ (a point-like membrane) and $\textbf{u}_{1/3}$ (the Myers dielectric sphere) and a saddle point at $\textbf{u}_{1/6}$ (unstable static sphere):
\begin{IEEEeqnarray}{c}
V_{\text{eff}}\left(0\right) = V_{\text{eff}}\left(\frac{1}{3}\right) = 0,\quad V_{\text{eff}}\left(\frac{1}{6}\right) = \frac{2\pi T \mu^{4}}{6^4}. \qquad \label{So3Extrema2}
\end{IEEEeqnarray}
\indent The radial stability analysis of the solution \eqref{StaticAnsatz} around the extremal points \eqref{So3Extrema1} was carried out in \cite{AxenidesFloratosLinardopoulos17a} where it was shown that $\textbf{u}_0$ and $\textbf{u}_{1/3}$ are radially stable, whereas $\textbf{u}_{1/6}$ is a radially unstable point.
\section[Leading-order perturbations]{Leading-order perturbations \label{Section:LeadingOrderPT}}
\noindent The study of angular perturbations around the spherically symmetric extrema \eqref{So3Extrema1} was performed in \cite{AxenidesFloratosLinardopoulos17b}. Below we revisit the main points of the analysis and restate its main results by supplying certain new details. \\
\indent The static ansatz $x_i\left(\tau\right) = \mu u_0 e_i$ is constructed from the $SO(3)$ extrema $u_0 = \{0,1/6,1/3\}$. It satisfies \eqref{xEquation1} for $y_k = 0$,
\begin{IEEEeqnarray}{ll}
\ddot{x}_i = &\left\{\left\{x_i,x_k\right\},x_k\right\} - \frac{\mu^2}{9}\,x_i + \frac{\mu}{2}\epsilon_{ikl}\left\{x_k,x_l\right\} \qquad \label{xEquation2}
\end{IEEEeqnarray}
and corresponds to the initial conditions $x_i(0) = \mu x_i^0$ and $\dot{x}_i(0) = 0$, at $\tau=0$. The perturbative analysis which was performed in the paper \cite{AxenidesFloratosLinardopoulos17b} is based on the work of Lyapunov (details can be found in many standard textbooks, e.g.\ \cite{Lefschetz57}). Consider an infinitesimal variation of the initial conditions (for $i = 1, 2, 3$)
\begin{IEEEeqnarray}{l}
x_i(0) = \mu\left(x_i^0 + \varepsilon \, \delta x_i(0)\right), \quad \dot{x}_i(0) = \mu \varepsilon \delta \dot{x}_i(0). \qquad \label{InitialConditionPerturbations1}
\end{IEEEeqnarray}
The perturbation of the initial conditions \eqref{InitialConditionPerturbations1} induces an $\varepsilon$-dependence in the solution $x_i(\tau,\varepsilon)$. In general, $x_i(\tau,\varepsilon)$ affords a series expansion in $\varepsilon$
\begin{IEEEeqnarray}{l}
x_i = \mu\big(x_i^0 + \sum_{n = 1}^{\infty}\varepsilon^n\delta x_i^{(n)}\big), \quad i = 1, 2, 3. \qquad \label{AngularPerturbations1}
\end{IEEEeqnarray}
Inserting the perturbation \eqref{AngularPerturbations1} into the equations of motion \eqref{xEquation2}, we obtain to leading order $\varepsilon$
\begin{IEEEeqnarray}{ll}
\delta\ddot{x}_i &= \left\{\left\{\delta x_i,x_j^0\right\},x_j^0\right\} + \left\{\left\{x_i^0,\delta x_j\right\},x_j^0\right\} + \nonumber \\
& + \left\{\left\{x_i^0,x_j^0\right\},\delta x_j\right\} - \frac{1}{9}\,\delta x_i + \epsilon_{ijk}\left\{\delta x_j,x_k^0\right\}, \qquad \label{AngularPerturbations2}
\end{IEEEeqnarray}
where, from now on, we switch to the dimensionless time $t \equiv \mu\tau$. For simplicity we have also set $\delta x_i^{(1)} \equiv \delta x_i$. The Gauss law constraint \eqref{GaussLaw1} becomes:
\begin{IEEEeqnarray}{c}
\left\{\delta\dot{x}_i, x_i^0\right\} = 0. \label{GaussLaw2}
\end{IEEEeqnarray}
It can be shown \cite{AxenidesFloratosLinardopoulos17b} that if \eqref{GaussLaw2} is satisfied at $t = 0$, then the perturbation equations \eqref{AngularPerturbations2} guarantee its validity at all times $t$. Following the works \cite{AxenidesFloratosPerivolaropoulos01}, we expand $\delta x$ in spherical harmonics:
\begin{IEEEeqnarray}{ll}
\delta x_i = \sum_{j,m}\eta_i^{jm}\left(t\right) Y_{jm}\left(\theta,\phi\right), \qquad \label{AngularPerturbations3}
\end{IEEEeqnarray}
where $\delta x$ must be real, so that the fluctuation modes $\eta_i^{jm}\left(t\right)$ satisfy
\begin{IEEEeqnarray}{ll}
\eta_i^{jm*}\left(t\right) = (-1)^{m} \eta_i^{j(-m)}\left(t\right). \label{RealityCondition}
\end{IEEEeqnarray}
By using the property of spherical harmonics,
\begin{IEEEeqnarray}{ll}
\left\{e_i,Y_{jm}\left(\theta,\phi\right)\right\} &= -i \hat{J}_i^{\,(j)} Y_{jm}\left(\theta,\phi\right) = \nonumber \\
& = -i \sum_{m'}\left(J_i\right)_{m'm}^{(j)} Y_{jm'}\left(\theta,\phi\right), \qquad \label{SphericalHarmonics1}
\end{IEEEeqnarray}
where $\hat{J}_i^{(j)}$ is the spin-$j$ angular momentum operator and $\left(J_i\right)^{(j)}_{mm'}$ the $2j+1$ dimensional (spin-$j$) matrix representation of $\mathfrak{su}\left(2\right)$, we can show that the fluctuation modes $\eta_i$ satisfy the following equations of motion:
\begin{IEEEeqnarray}{c}
\ddot\eta_i + \omega_3^2 \eta_i = u_0^2 T_{ik} \eta_k + u_0 Q_{ik} \eta_k. \qquad \label{AngularPerturbations4}
\end{IEEEeqnarray}
In \eqref{AngularPerturbations4}, we have omitted the indices $j,m$, we have summed the repeated indices and have used the definitions
\begin{IEEEeqnarray}{ll}
\omega_3^2 \equiv u_0^2 \textbf{J}^2 + \frac{1}{9}, \qquad & T_{ik} \equiv J_i J_k - 2i\epsilon_{ikl}J_l \qquad \label{Definitions1} \\
& Q_{ik} \equiv i\epsilon_{ikl}J_l, \qquad \label{Definitions2}
\end{IEEEeqnarray}
where $\textbf{J}^2 = j(j+1)$. \eqref{AngularPerturbations4} can be written in a compact form as follows
\begin{IEEEeqnarray}{c}
\ddot{H} + \mathcal{K} \cdot H = 0, \qquad \mathcal{K} \equiv \omega_3^2 - u_0^2 T - u_0 Q, \qquad \label{AngularPerturbations5}
\end{IEEEeqnarray}
where $H$, $Q$ and $T$ refer to the $3\times(2j + 1)$ dimensional representations of $\eta_i^{jm}$, $Q_{ik}$ and $T_{ik}$:
\begin{IEEEeqnarray}{l}
H = \begin{pmatrix} \eta_x^{jm} \\ \eta_y^{jm} \\ \eta_z^{jm} \end{pmatrix}, \quad Q = i\begin{pmatrix} 0 & J_z & -J_y \\ -J_z & 0 & J_x \\ J_y & -J_x & 0 \end{pmatrix} \\[6pt]
T = \begin{pmatrix} J_x^2 & J_x J_y - 2i J_z & J_x J_z + 2i J_y \\ J_y J_x + 2i J_z & J_y^2 & J_y J_z - 2i J_x \\ J_z J_x - 2i J_y & J_z J_x + 2i J_x & J_z^2 \end{pmatrix}. \qquad
\end{IEEEeqnarray}
The $2j + 1$ and the $3\times(2j + 1)$ dimensional representations \eqref{AngularPerturbations4}, \eqref{AngularPerturbations5} are in many ways inequivalent. Many important details of the analysis that follows depend crucially on which of the two representations is being used.
\subsection[Eigenvalues]{Eigenvalues}
\noindent In order to solve \eqref{AngularPerturbations5} we set $H = e^{i\lambda t}\xi$. We are led to the following eigenvalue problem in the $3\times(2j + 1)$ dimensional space:
\begin{IEEEeqnarray}{c}
\left(-\lambda^2 + \omega_3^2 - u_0^2 \, T - u_0 \, Q\right) \cdot \xi = 0, \quad \xi \equiv (\xi_i^{jm}). \qquad \label{AngularPerturbations6}
\end{IEEEeqnarray}
We further introduce the $(2j+1)\times(2j+1)$ matrices
\begin{IEEEeqnarray}{ll}
P_{ik} \equiv &\frac{1}{j\left(j+1\right)} \, J_i J_k, \label{ProjectorDefinitionP} \\
R_{ik}^{\pm} \equiv &\frac{1}{2j +1} \bigg[\frac{1}{2}\left(2j+1\mp1\right)\cdot \left(\delta_{ik}\times I - P_{ik}\right) \pm \nonumber \\
& \hspace{1.5cm} \pm \left(\delta_{ik}\times I - Q_{ik}\right)\bigg], \qquad
\end{IEEEeqnarray}
which form a complete and orthogonal set of projection operators (i.e.\ idempotent and Hermitian). In the $3\times(2j+1)$ dimensional space,
\begin{IEEEeqnarray}{l}
P^2 = P = P^\dag, \quad R_{\pm}^2 = R_{\pm} = R_{\pm}^\dag, \label{AngularEigenvalueProblem1a} \\
P \cdot R^{\pm} = R^{+} \cdot R^{-} = 0, \label{AngularEigenvalueProblem1b} \\
P + R^{+} + R^{-} = I, \qquad \ \label{AngularEigenvalueProblem1c}
\end{IEEEeqnarray}
where $I$ is the $3(2j+1)\times3(2j+1)$ unit matrix. It turns out that the matrices $T$ and $Q$ are also Hermitian and can be expressed in terms of the projectors $P$, $R_{\pm}$:
\begin{IEEEeqnarray}{l}
T = \left[j\left(j+1\right) - 2\right]P + 2jR_+ - 2\left(j+1\right)R_-, \qquad \label{AngularEigenvalueProblem2a} \\
Q = P - j R_+ + \left(j+1\right)R_-. \label{AngularEigenvalueProblem2b}
\end{IEEEeqnarray}
The advantage of this decomposition is that we can immediately read the eigenvalues of the matrices $T$ and $Q$ (along with their degeneracies) in each of the projective spaces $P,R_{\pm}$. Plugging \eqref{AngularEigenvalueProblem2a}--\eqref{AngularEigenvalueProblem2b} into \eqref{AngularPerturbations6} we are led to
\begin{IEEEeqnarray}{l}
\left(\omega_3^2 - \lambda^2\right)\xi = \Bigg[\left(u_0^2\left[j\left(j+1\right) - 2\right] + u_0\right)P + \nonumber \\
+ j u_0 \cdot \left(2u_0 - 1\right)R_+ - \left(j+1\right)u_0\left(2u_0 - 1\right)R_-\Bigg]\xi, \qquad \ \label{AngularPerturbations7}
\end{IEEEeqnarray}
which we then multiply with the projectors $P$ and $R_{\pm}$ in order to obtain the eigenvalues $\lambda$:
\begin{IEEEeqnarray}{l}
\lambda_P^2 = 2(u_0 - \frac{1}{3})(u_0 - \frac{1}{6}), \qquad \label{Eigenfrequency1a} \\
\lambda_+^2 = j\left(j-1\right)u_0^2 + j u_0 + \frac{1}{9}, \qquad \label{Eigenfrequency1b} \\
\lambda_-^2 = \left(j+1\right)\left(j+2\right)u_0^2 - \left(j+1\right)u_0 + \frac{1}{9}. \qquad \label{Eigenfrequency1c}
\end{IEEEeqnarray}
\indent The degeneracies of the eigenvalues $\lambda_{P,\pm}^2$ are equal to the dimensionalities of the corresponding projectors $P$, $R_{\pm}$, i.e.\ $d_+ = 2j+3$, $d_P = 2j+1$, $d_- = 2j-1$. For each of the critical points in \eqref{So3Extrema1} we find:
\begin{IEEEeqnarray}{ll}
\textbf{u}_0: \quad &\lambda_P^2 = \lambda_{\pm}^2 = \frac{1}{9} \qquad \label{Eigenfrequency2a} \\
\textbf{u}_{1/6}: \quad &\lambda_P^2 = 0, \quad \lambda_+^2 =\frac{1}{36}\left(j + 1\right)\left(j + 4\right), \nonumber \\
& \lambda_-^2 = \frac{j\left(j-3\right)}{36} \qquad \label{Eigenfrequency2b} \\
\textbf{u}_{1/3}: \quad &\lambda_P^2 = 0, \quad \lambda_+^2 =\frac{1}{36}\left(j + 1\right)^2, \quad \lambda_-^2 = \frac{j^2}{9}, \qquad \ \label{Eigenfrequency2c}
\end{IEEEeqnarray}
which coincide with the eigenvalues that were found in \cite{DasguptaJabbariRaamsdonk02} for the BMN matrix model. Note that only the $R_{-}$ sector of $\textbf{u}_{1/6}$ is unstable for $j=1,2$ ($\lambda_-$ is purely imaginary). These instabilities are related to the existence of a separatrix in the corresponding phase diagram (see e.g.\ \cite{AxenidesFloratosLinardopoulos17a}). \\
\indent The general solution of \eqref{AngularPerturbations5} can be written in the $3\times(2j+1)$ dimensional space as follows:
\begin{IEEEeqnarray}{ll}
H\left(t\right) = &e^{i\lambda_{P}t}\xi_P + e^{-i\lambda_{P}t}\tilde{\xi}_P + e^{i\lambda_{+}t}\xi_+ + e^{-i\lambda_{+}t}\tilde{\xi}_+ + \nonumber \\ & + e^{i\lambda_{-}t}\xi_- + e^{-i\lambda_{-}t}\tilde{\xi}_-, \label{GeneralSolution1}
\end{IEEEeqnarray}
where $\xi_{A}, \ \tilde{\xi}_{A}$, $A \equiv \{P,\pm\}$ are generic $3\times(2j+1)$ dimensional vectors of the subspaces $P$, $R_{\pm}$. These are naturally expressed as linear combinations of the corresponding eigenvectors $\vert P,\pm \rangle$. They are determined by the initial values of $H$, $\dot{H}$ at $t=0$ and the leading order of the Gauss law constraint (see \eqref{ConstraintFirstOrder} below).
\subsection[Eigenvectors]{Eigenvectors}
\noindent Calculating the square of the matrix $Q$ (defined in \eqref{Definitions2} above) by using the decomposition \eqref{AngularEigenvalueProblem2a}--\eqref{AngularEigenvalueProblem2b}, we find that the projection operators $P$ and $R_{\pm}$ can be expressed in terms of $Q$ as follows:
\begin{IEEEeqnarray}{ll}
P = & I + \frac{Q - Q^2}{j(j+1)}, \quad
R_{\pm} = \frac{1}{2j +1} \bigg[\frac{1}{2}\left(2j+1\mp1\right)\cdot \nonumber \\
& \cdot \frac{Q^2 - Q}{j(j+1)} \pm \left(I - Q\right)\bigg]. \qquad \label{AngularEigenvalueProblem3}
\end{IEEEeqnarray}
This implies that the eigenvectors of $P$ and $R_{\pm}$ are fully determined by those of $Q$. \\
\indent Before going on to derive the precise expressions of these eigenvectors, let us note that $Q$ is just the spin-orbit coupling operator of the orbital angular momentum $L = 1$ with the spin angular momentum $J = j$. To see why this is so, consider the so-called adjoint representation of $SU(2)$ in which the three components of the orbital angular momentum $L = 1$ read:
\begin{IEEEeqnarray}{l}
\left(L_i\right)_{kl} = i\epsilon_{ilk} \quad \Rightarrow \quad
L_x = \begin{pmatrix}
0 & 0 & 0\\
0 & 0 & -i\\
0 & i &0
\end{pmatrix}, \nonumber \\[6pt]
L_y = \begin{pmatrix}
0 & 0 & i\\
0 & 0 & 0\\
-i & 0 & 0
\end{pmatrix}, \quad
L_z = \begin{pmatrix}
0 & -i & 0\\
i & 0 & 0\\
0 & 0 & 0
\end{pmatrix} \label{OrbitalAngularMomentum}
\end{IEEEeqnarray}
and $Q$ indeed takes the form of spin-orbit coupling,
\begin{IEEEeqnarray}{c}
Q_{ik} = \left(L_l\right)_{ki} J_l \quad \Leftrightarrow \quad Q = -L_i \otimes J_i, \label{SpinOrbitCoupling1}
\end{IEEEeqnarray}
or, in terms of the raising and lowering operators $L_{\pm} \equiv L_x \pm i L_y$ and $J_{\pm} \equiv J_x \pm i J_y$,
\begin{IEEEeqnarray}{c}
Q = -\frac{1}{2}\left(L_ + \otimes J_-\right) - \frac{1}{2}\left(L_-\otimes J_+\right) - L_z\otimes J_z. \qquad \label{SpinOrbitCoupling2}
\end{IEEEeqnarray}
The three components of the total angular momentum $\textbf{J}_{\scaleto{T}{5pt}} = \textbf{L} + \textbf{J}$, for $J = j$, $L = 1$ are given by
\begin{IEEEeqnarray}{c}
J_{\scaleto{T}{5pt}}^{i} = J_{i}\otimes I_{3} + I_{2j+1}\otimes L_{i}
\end{IEEEeqnarray}
and explicitly,
\begin{IEEEeqnarray}{c}
J_{\scaleto{T}{5pt}}^x = \begin{pmatrix}
J_x & 0 & 0\\
0 & J_x & -i I\\
0 & i I & J_x
\end{pmatrix}, \qquad
J_{\scaleto{T}{5pt}}^y = \begin{pmatrix}
J_y & 0 & i I\\
0 & J_y & 0\\
-i I & 0 & J_y
\end{pmatrix}, \nonumber \\[6pt]
J_{\scaleto{T}{5pt}}^z = \begin{pmatrix}
J_z & -i I & 0\\
i I & J_z & 0\\
0 & 0 & J_z
\end{pmatrix}, \qquad
\end{IEEEeqnarray}
while the square of the total angular momentum is
\begin{IEEEeqnarray}{c}
\textbf{J}_{\scaleto{T}{5pt}}^2 = \left(j(j+1)+2\right)I_{3(2j+1)}-2Q, \label{TotalAngularMomentum1}
\end{IEEEeqnarray}
so that its eigenstates,
\begin{IEEEeqnarray}{l}
\vert j+1,m;j,1\rangle, \quad \vert j,m;j,1\rangle, \quad \vert j-1,m;j,1\rangle, \qquad
\end{IEEEeqnarray}
will obviously also diagonalize the spin-orbit coupling operator $Q$. \\
\indent By directly diagonalizing the matrix $Q$ in \eqref{SpinOrbitCoupling2}, or equivalently, by means of the standard Clebsch-Gordan analysis (see e.g.\ \cite{Rose57}) we find
\begin{widetext}
\vspace{-.3cm}\begin{IEEEeqnarray}{l}
\vert j+1,m;j,1\rangle = \sqrt{\frac{\left(j+m\right)\left(j+m+1\right)}{2\left(j+1\right)\left(2j+1\right)}}\cdot\vert 1,1\rangle\vert j,m-1\rangle + \sqrt{\frac{\left(j+1\right)^2-m^2}{(j+1)\left(2j+1\right)}}\cdot\vert 1,0\rangle\vert j,m\rangle + \nonumber \\
\hspace{8.8cm} + \sqrt{\frac{\left(j-m\right)\left(j-m+1\right)}{2(j+1)\left(2j+1\right)}}\cdot\vert 1,-1\rangle\vert j,m+1\rangle, \qquad \label{TotalAngularMomentum2a} \\
\vert j,m;j,1\rangle = -\sqrt{\frac{\left(j+m\right)\left(j-m+1\right)}{2j\left(j+1\right)}}\cdot\vert1,1\rangle\vert j,m-1\rangle + \frac{m}{\sqrt{j(j+1)}}\cdot\vert1,0\rangle\vert j,m\rangle + \nonumber \\
\hspace{8.8cm} + \sqrt{\frac{\left(j-m\right)\left(j+m+1\right)}{2j\left(j+1\right)}}\cdot\vert1,-1\rangle\vert j,m+1\rangle, \qquad \label{TotalAngularMomentum2b} \\
\vert j-1,m;j,1\rangle = \sqrt{\frac{\left(j-m\right)\left(j-m+1\right)}{2j\left(2j+1\right)}}\cdot\vert 1,1\rangle\vert j,m-1\rangle - \sqrt{\frac{j^2-m^2}{j\left(2j+1\right)}}\cdot\vert 1,0\rangle\vert j,m\rangle + \nonumber \\
\hspace{8.8cm} + \sqrt{\frac{\left(j+m\right)\left(j+m+1\right)}{2j\left(2j+1\right)}}\cdot\vert 1,-1\rangle\vert j,m+1\rangle. \qquad\label{TotalAngularMomentum2c}
\end{IEEEeqnarray}
\end{widetext}
Acting with the spin-orbit coupling operator $Q$ on the Clebsch-Gordan system \eqref{TotalAngularMomentum2a}--\eqref{TotalAngularMomentum2c}, we obtain the following eigenvalues:
\begin{IEEEeqnarray}{l}
Q\cdot\vert j + 1,m;j,1\rangle = -j \,\vert j + 1,m;j,1\rangle, \qquad \\
Q\cdot\vert j,m;j,1\rangle = \vert j,m;j,1\rangle, \qquad \\
Q\cdot\vert j - 1,m;j,1\rangle = \left(j+1\right)\vert j - 1,m;j,1\rangle. \qquad
\end{IEEEeqnarray}
Furthermore, by plugging the eigenstates \eqref{TotalAngularMomentum2a}--\eqref{TotalAngularMomentum2c} into the expressions \eqref{AngularEigenvalueProblem3} that give the projectors $P$, $R_{\pm}$ in terms of the spin-orbit coupling operator $Q$, we find that the eigenstates $\vert j,m;j,1\rangle$ span the subspace of the projector $P$, while the eigenstates $\vert j \pm 1,m;j,1\rangle$ span the subspaces of the projectors $R_{\pm}$, i.e.\
\begin{IEEEeqnarray}{c}
\vert P\rangle = \vert j,m;j,1\rangle, \qquad \vert \pm \rangle = \vert j \pm 1,m;j,1\rangle. \qquad
\end{IEEEeqnarray}
\subsection[Gauss law contraint (LO)]{Gauss law contraint (LO)}
\noindent Plugging the solution \eqref{StaticAnsatz} and the perturbative expansion \eqref{AngularPerturbations3} into the Gauss law constraint \eqref{GaussLaw2} we obtain, to LO in perturbation theory
\begin{IEEEeqnarray}{c}
\sum_{j,m}\dot{\eta}_i^{jm}\left\{Y_{jm}, e_i\right\} = \sum_{j,m,m'}\dot{\eta}_i^{jm} \left(J_i\right)_{m'm}^{(j)} Y_{jm'} = 0 \Rightarrow \nonumber \\
\Rightarrow \sum_{m}\dot{\eta}_i^{jm} \left(J_i\right)_{m'm}^{(j)} = 0,
\end{IEEEeqnarray}
since the spherical harmonics $Y_{jm}$ form an orthonormal basis. Multiplying with $J_k^{(j)}$ and using the definition \eqref{ProjectorDefinitionP} of the projector $P$ we find,
\begin{IEEEeqnarray}{c}
\sum_{m, m'}\dot{\eta}_i^{jm} \left(J_k\right)_{m''m'}^{(j)}\left(J_i\right)_{m'm}^{(j)} = 0 \Rightarrow \nonumber \\
\Rightarrow \sum_{m'} \left(P_{ik}\right)_{mm'}^{(j)} \dot{\eta}_k^{jm'} = 0. \label{ConstraintFirstOrder}
\end{IEEEeqnarray}
\eqref{ConstraintFirstOrder} constrains the generic form \eqref{GeneralSolution1} of the LO perturbations $\eta_i^{jm}$, which in principle span all the three sectors $P, \ R_{\pm}$. \\
\indent Because of the Gauss law constraint \eqref{ConstraintFirstOrder}, $\eta_i^{jm}$ is forced to live exclusively inside the sectors $R_{\pm}$ at all the critical points \eqref{Eigenfrequency2a}--\eqref{Eigenfrequency2c}. However, it receives an additional contribution from the zero eigenvalue eigenstate $\xi_P$ of the $P$-sector at the critical points $\textbf{u}_{1/6}$ and $\textbf{u}_{1/3}$. \\
\section[Higher-order perturbations]{Higher-order perturbations \label{Section:HigherOrderPT}}
\noindent Let us now study the perturbative expansion \eqref{AngularPerturbations1} beyond the leading order. The initial conditions for the perturbations $\delta x_i^{(n)}$ and their derivatives are determined from the initial conditions of the complete solution \eqref{AngularPerturbations1}. They all vanish at $t = 0$,
\begin{IEEEeqnarray}{ll}
\delta x_i^{(n)}(0) = \delta \dot{x}_i^{(n)}(0) = 0, \qquad n = 2,3,\ldots, \label{InitialConditionPerturbations2}
\end{IEEEeqnarray}
unless $n = 0,1$. To obtain the equations of motion for the perturbations, we plug the series \eqref{AngularPerturbations1} into the $SO(3)$ equations of motion \eqref{xEquation2}. We are led to
\begin{widetext}
\begin{IEEEeqnarray}{ll}
\delta\ddot{x}_i^{(n)} = &\{\{\delta x_i^{(n)},x^{(0)}_k\},x^{(0)}_k\} + \{\{x^{(0)}_i,\delta x_k^{(n)}\},x^{(0)}_k\} + \{\{x^{(0)}_i,x^{(0)}_k\},\delta x_k^{(n)}\} - \frac{1}{9}\delta x_i^{(n)} + \epsilon_{ikl}\{x^{(0)}_k,\delta x_l^{(n)}\} + \nonumber \\
& + \sum_{p = 1}^{n-1}\Bigg[\{\{x^{(0)}_i,\delta x_k^{(n-p)}\},\delta x_k^{(p)}\} + \{\{\delta x_i^{(n-p)}, x^{(0)}_k\},\delta x_k^{(p)}\} + \{\{\delta x_i^{(n-p)},\delta x_k^{(p)}\},x^{(0)}_k\} + \nonumber \\
& + \frac{1}{2}\epsilon_{ikl}\{\delta x_k^{(n-p)},\delta x_l^{(p)}\} \Bigg] + \sum_{p = 1}^{n-1} \sum_{q = 1}^{p-1}\{\{\delta x_i^{(n-p)},\delta x_k^{(p-q)}\},\delta x_k^{(q)}\}. \qquad \label{HigherOrderPerturbations2}
\end{IEEEeqnarray}
\end{widetext}
As before, we proceed to expand the perturbations in spherical harmonics,
\begin{IEEEeqnarray}{ll}
\delta x_i^{(n)} = \sum_{j,m}\eta_i^{njm}\left(t\right) Y_{jm}\left(\theta,\phi\right), \ &\eta_i^{njm}\left(0\right) = 0, \qquad \ \label{AngularPerturbations8}
\end{IEEEeqnarray}
which, in addition to the reality condition \eqref{RealityCondition} and the property \eqref{SphericalHarmonics1}, obey
\begin{IEEEeqnarray}{c}
\left\{Y_{\alpha}\left(\theta,\phi\right),Y_{\beta}\left(\theta,\phi\right)\right\} = f_{\alpha\beta}^{\gamma} Y_{\gamma}\left(\theta,\phi\right), \label{SphericalHarmonics2} \qquad
\end{IEEEeqnarray}
where $f_{\alpha\beta}^{\gamma}$ are the structure constants of the area-preserving symmetry of the Hamiltonian \eqref{ppWaveHamiltonian} and $\alpha \equiv jm$, $\beta \equiv j'm'$, $\gamma \equiv j''m''$. The structure constants $f_{\alpha\beta}^{\gamma}$ can be computed by means of a closed formula that is valid for all values of the spin quantum numbers $\alpha$, $\beta$, $\gamma$. The expressions for the structure constants $f_{1m,\beta}^{\gamma}$ and $f_{2m,\beta}^{\gamma}$ can be found in appendix \ref{Appendix:StructureConstants}.
\subsection[Next-to-leading order perturbations]{Next-to-leading order perturbations}
\noindent For $n = 2$, it is easy to show that the higher-order perturbation equations \eqref{HigherOrderPerturbations2} become:
\begin{widetext}
\begin{IEEEeqnarray}{ll}
\delta\ddot{x}_i^{(2)} = &\{\{\delta x_i^{(2)},x^{(0)}_k\},x^{(0)}_k\} + \{\{x^{(0)}_i,\delta x_k^{(2)}\},x^{(0)}_k\} + \{\{x^{(0)}_i,x^{(0)}_k\},\delta x_k^{(2)}\} - \frac{1}{9}\delta x_i^{(2)} + \epsilon_{ikl}\{x^{(0)}_k,\delta x_l^{(2)}\} + \nonumber \\
& + \Bigg[\{\{x^{(0)}_i,\delta x_k^{(1)}\},\delta x_k^{(1)}\} + \{\{\delta x_i^{(1)}, x^{(0)}_k\},\delta x_k^{(1)}\} + \{\{\delta x_i^{(1)},\delta x_k^{(1)}\},x^{(0)}_k\} + \frac{1}{2}\epsilon_{ikl}\{\delta x_k^{(1)},\delta x_l^{(1)}\}\Bigg], \qquad \label{HigherOrderPerturbations3}
\end{IEEEeqnarray}
\end{widetext}
so that by using \eqref{AngularPerturbations8}, \eqref{SphericalHarmonics1}, \eqref{SphericalHarmonics2}, we arrive at
\begin{IEEEeqnarray}{l}
\ddot\eta_i^{(2)} + \omega_3^2 \eta_i^{(2)} = u_0^2 T_{ik} \eta_k^{(2)} + u_0 Q_{ik} \eta_k^{(2)} + F_i^{(2)}. \qquad \label{AngularPerturbations9}
\end{IEEEeqnarray}
For simplicity, we have suppressed the spin indices $j,m$ and have made all time dependence implicit in \eqref{AngularPerturbations9}. Notice that the $n = 2$ equations \eqref{AngularPerturbations9} are just the $n=1$ equations \eqref{AngularPerturbations4}, driven by the forcing term $F_i^{2jm}\left(t\right)$. The latter can be written as a bilinear form
\begin{IEEEeqnarray}{l}
F_i^{2\gamma}\left(t\right) = \eta_{k}^{1\alpha} K_{ikl;\alpha\beta}^{\gamma} \eta_{l}^{1\beta},
\end{IEEEeqnarray}
where once more, we denote $\alpha \equiv jm$, $\beta \equiv j'm'$, $\gamma \equiv j''m''$ and have omitted the sums on the spatial indices $k,l$ and the spin indices $\alpha,\beta$. The matrix of the bilinear form $K$ is given by
\begin{IEEEeqnarray}{ll}
K_{ikl;\alpha\beta}^{\gamma} = &\left(\mathcal{J}_a\right)^{(j)}_{\dot{m}m} f_{j\dot{m},\beta}^{\gamma}\left(\epsilon_{bak}\epsilon_{bil} + \epsilon_{bai}\epsilon_{bkl}\right) + \nonumber \\
& + \frac{1}{2}\epsilon_{ikl}f_{\alpha\beta}^{\gamma}, \hspace{.7cm} \label{ForcingOrder2}
\end{IEEEeqnarray}
where $\mathcal{J}_a \equiv - i u_0 J_a$ and we have omitted the dependence of $\eta_i^{1jm}\left(t\right)$ on the dimensionless time $t \equiv \mu\tau$. As usual, we have also omitted the sums in the spatial indices $a,b$ and the spin index $\dot{m}$ in \eqref{ForcingOrder2}. In compact form, \eqref{AngularPerturbations9} can be written as
\begin{IEEEeqnarray}{c}
\ddot{H}^{(2)} + \mathcal{K} \cdot H^{(2)} = F^{(2)}, \quad F^{(2)} \equiv H^{(1)}\,K \, H^{(1)}. \qquad \label{AngularPerturbations10}
\end{IEEEeqnarray}
\indent Let us now discuss the role of the LO unstable modes. The presence of unstable LO modes (i.e.\ having $\lambda_{-}^2<0$ in \eqref{Eigenfrequency2b}) in the forcing \eqref{ForcingOrder2} can induce further unstable modes at the NLO, for any spin $j$. The form of the structure constants $f_{1m,\beta}^{\gamma}$ and $f_{2m,\beta}^{\gamma}$ (cf.\ \eqref{StructureConstants2}--\eqref{StructureConstants5} in appendix \ref{Appendix:StructureConstants}), implies that the coupling of an unstable LO mode with $j=1$ to a stable LO mode $j'$ can destabilize all the $j''=j'$ modes at NLO order. Similarly, the coupling of an unstable LO mode with $j=2$ to a stable LO mode $j'$ can destabilize all the $j''=j'\pm 1$ modes at NLO order. The cascade of perturbative instabilities extends to the higher perturbative orders, so that all the modes at any given perturbative order $n$ can become unstable because of the LO instabilities at $j=1,2$. \\
\indent Another possible source of instabilities shows up whenever the frequency of the external forcing $F_i^{2\gamma}\left(t\right) $ in \eqref{ForcingOrder2} matches one of the system's natural eigenfrequencies \eqref{Eigenfrequency2a}--\eqref{Eigenfrequency2c}. These resonances are not restricted to the unstable critical point $\textbf{u}_{1/6}$ but they may also destabilize the stable critical point $\textbf{u}_{1/3}$ (i.e.\ the Myers sphere). For example, a resonance will occur when a zero mode in the $P$ sector couples to a mode in the $R_{\pm}$ sectors in such a way that the resulting forcing \eqref{ForcingOrder2} contains at least one of the system's natural eigenfrequencies \eqref{Eigenfrequency2a}--\eqref{Eigenfrequency2c}. The above discussion has to take into consideration the LO and NLO Gauss law constraints which impose the symmetries to the solution of the equations.
\subsection[Gauss law contraint (NLO)]{Gauss law contraint (NLO)}
\noindent The LO constraint equation \eqref{ConstraintFirstOrder} implies that the initial conditions for the velocity of any mode $\eta_i^{(1)}$ must be orthogonal to the $P$ sector and thus it should be a linear combination of the eigenvectors of the other two sectors $R_{\pm}$. The NLO Gauss law constraint \eqref{GaussLaw2} reads
\begin{equation}
\left\{\dot{x}^{(2)}_i,x_i^{(0)} \right\} + \left\{\dot{x}^{(1)}_i,x_i^{(1)} \right\} = 0,
\end{equation}
since $\dot{x}^{(0)}_i = 0$. Substituting the values of $x_i^{(0)}$, $\dot{x}^{(n)}_i$ from \eqref{StaticAnsatz}, \eqref{AngularPerturbations3} we obtain
\begin{IEEEeqnarray}{l}
u_0 \dot{\eta}_i^{2jm} \left\{Y_{jm},e_i \right\} + \dot{\eta}_i^{1jm}\eta_i^{1j^\prime m^\prime} \left\{Y_{jm},Y_{j^\prime m^\prime} \right\} = 0, \qquad
\end{IEEEeqnarray}
where the sums over all the repeated indices have been omitted as usual. Calculating the brackets by using \eqref{SphericalHarmonics1} and \eqref{SphericalHarmonics2} it follows that
\begin{equation}
i \, u_0 \,\dot{\eta}_i^{2jm} \left(J_i\right)^{(j)}_{m^\prime m} Y_{jm^\prime} + \dot{\eta}_i^{1jm} \eta_i^{1j^\prime m^\prime} f_{jm,j^\prime m^\prime}^{j^{\prime\prime}m^{\prime\prime}} Y_{j^{\prime\prime}m^{\prime\prime}} = 0,
\end{equation}
which gives
\begin{IEEEeqnarray}{c}
i \, u_0 \, \dot{\eta}_i^{2j^{\prime\prime}m} \left(J_i\right)^{(j^{\prime\prime})}_{m^{\prime\prime}m} + \dot{\eta}_i^{1jm} \eta_i^{1j^\prime m^\prime} f_{jm,j^\prime m^\prime}^{j^{\prime\prime}m^{\prime\prime}} = 0, \qquad \label{ConstraintSecondOrder}
\end{IEEEeqnarray}
after factoring out the spherical harmonics. This equation implies that the initial velocity $\dot{\eta}_i^{(2)}$ must be a linear combination of the $R_{\pm}$ sectors and a component that belongs to the $P$ sector and is given by the LO forcing term that appears in the constraint. The latter depends on the coupling of the initial velocity $\dot{\eta}_i^{(1)}$ and the values of $\eta_i^{(1)}$ for different spins $j$.
\subsection[General solution]{General solution}
\noindent The general solution of the NLO perturbation equations \eqref{AngularPerturbations10} can be written as the sum of the general solution of the homogeneous equation \eqref{AngularPerturbations5} and a special solution of the full equation \eqref{AngularPerturbations10}:
\begin{IEEEeqnarray}{c}
H^{(2)}\left(t\right) = H^{(2)}_{\text{h}}\left(t\right) + H^{(2)}_{\text{s}}\left(t\right).
\end{IEEEeqnarray}
We have already shown that the general solution $H^{(2)}_{\text{h}}\left(t\right)$ of the homogeneous equation \eqref{AngularPerturbations5} takes the form \eqref{GeneralSolution1}. By writing $\mathcal{K} \equiv \omega_3^2 - u_0^2 T - u_0 Q = \Omega_0^2$, it is easy to see that $H^{(2)}_{\text{h}}\left(t\right)$ can also be written as
\begin{IEEEeqnarray}{c}
H^{(2)}_{\text{h}}\left(t\right) = H^{(2)}_{\text{h}}\left(0\right)\cos\Omega_0 t + \dot{H}^{(2)}_{\text{h}}\left(0\right)\Omega_0^{-1}\sin\Omega_0 t. \qquad
\end{IEEEeqnarray}
Similarly, the special solution $H^{(2)}_{\text{s}}\left(t\right)$ takes the following form:
\begin{IEEEeqnarray}{ll}
H^{(2)}_{\text{s}}\left(t\right) &= \Omega_0^{-1}\sin\left(\Omega_0 t\right)\int_{0}^{t}ds\cos\left(\Omega_0 s\right)F^{(2)}\left(s\right) \nonumber \\
& - \Omega_0^{-1}\cos\left(\Omega_0 t\right)\int_{0}^{t}ds\sin\left(\Omega_0 s\right)F^{(2)}\left(s\right). \qquad \label{GeneralSolution2}
\end{IEEEeqnarray}
Interestingly, the structure of the complete solution \eqref{GeneralSolution2} remains the same at all higher perturbative orders $n = 2,3,\ldots$, while it turns out that the corresponding forcing terms $F^{(n)}(s)$ depend on the solutions of the previous orders.
\section[Example]{Example \label{Section:Example}}
\begin{figure}[h]
\begin{center}
\includegraphics[scale=0.4]{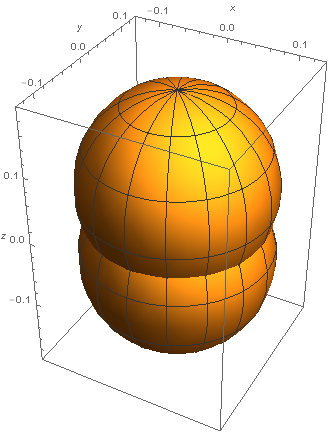}
\end{center}
\caption{Parametric plot of the forcing \eqref{Forcing1a}--\eqref{Forcing1c}.\label{Figure:ForcingPlot}}
\end{figure}

\noindent We now work out a simple example that demonstrates the instability cascade phenomenon. Let us compute $F_i^{2\gamma}$ at the point $u_0 = 1/6$ and spin $j'' = 3$, when only the LO mode $\xi^{2,0}_{-} \equiv \xi$ is turned on. Then \eqref{GeneralSolution1} becomes:
\begin{IEEEeqnarray}{c}
H\left(t\right) = e^{i\lambda_{-}t}\xi_{-}, \qquad \xi_{-} = \xi\cdot\vert - \rangle\Big|_{j=2, m=0}. \qquad
\end{IEEEeqnarray}
Computing the value \eqref{TotalAngularMomentum2c} of the eigenvector $\vert - \rangle$ for $j=2$, $m=0$ it is easy to show that only the following $j=2$ components of $\eta_i^{1\alpha}$ are nonzero:
\begin{IEEEeqnarray}{l}
\eta_x^{1,2,\pm1} = \pm\frac{\xi}{2} \, \sqrt{\frac{3}{5}} \cdot e^{t/(3\sqrt{2})}, \label{Solution1} \\
\eta_y^{1,2,\pm1} = -\frac{i\xi}{2}\,\sqrt{\frac{3}{5}} \cdot e^{t/(3\sqrt{2})}, \label{Solution2} \\
\eta_z^{1,2,0} = -\xi\,\sqrt{\frac{2}{5}} \cdot e^{t/(3\sqrt{2})}. \label{Solution3}
\end{IEEEeqnarray}
It can be shown that the solution \eqref{Solution1}--\eqref{Solution3} satisfies both the reality condition \eqref{RealityCondition} and the LO Gauss constraint \eqref{ConstraintFirstOrder}. Plugging \eqref{Solution1}--\eqref{Solution3} into the expression \eqref{ForcingOrder2} for the NLO forcing term, we obtain the following nonzero components:
\begin{IEEEeqnarray}{l}
F_x^{2,3,\pm 1} = \pm\frac{2\xi^2}{5} \sqrt{\frac{3}{7\pi}}\cdot e^{\sqrt{2}t/3}, \label{Forcing1a} \\
F_y^{2,3,\pm 1} = -\frac{2i\xi^2}{5} \sqrt{\frac{3}{7\pi}}\cdot e^{\sqrt{2}t/3}, \label{Forcing1b} \\
F_z^{2,3,0} = -\frac{6\xi^2}{5\sqrt{7\pi}} \cdot e^{\sqrt{2}t/3}, \qquad \label{Forcing1c}
\end{IEEEeqnarray}
by using the analytic expressions \eqref{StructureConstants3}--\eqref{StructureConstants5} for the structure constants $f_{\alpha\beta}^{\gamma}$ and the well-known angular momentum matrices $J_i$ (in the spin-$j$ representation). The parametric plot of the forcing \eqref{Forcing1a}--\eqref{Forcing1c} can be found in figure \ref{Figure:ForcingPlot}.
Because the only nonzero components of the LO modes are \eqref{Solution1}--\eqref{Solution3}, we only need the following structure constants
\begin{IEEEeqnarray}{l}
f_{2,0;2,1}^{3,1} = -f_{2,1;2,0}^{3,1} = -f_{2,0;2,-1}^{3,-1} = f_{2,-1;2,0}^{3,-1} = \nonumber \\
= -3i\sqrt{\frac{2}{7\pi}}, \qquad f_{2,1;2,-1}^{3,0} = -f_{2,-1;2,1}^{3,0} = \frac{6i}{\sqrt{7\pi}}. \qquad
\end{IEEEeqnarray}
\indent We next insert the expression for the forcing \eqref{Forcing1a}--\eqref{Forcing1c} into the $n=2$ perturbation equation \eqref{AngularPerturbations9} which we want to solve for the mode $\eta_i^{2,3,m}$ that was metastable for $n = 1$ (cf.\ equation \eqref{Eigenfrequency2b}). As usual, the general solution of \eqref{AngularPerturbations9} can be cast in the following form:
\begin{IEEEeqnarray}{l}
\eta_k^{2\gamma}\left(t\right) = \tilde{\eta}_{k}^{\gamma}\left(t\right) + \zeta_{k}^{\gamma} \, e^{\sqrt{2}t/3}, \quad k = 1,2,3, \qquad \label{GeneralSolution3}
\end{IEEEeqnarray}
where $\tilde{\eta}_{k}^{\gamma}\left(t\right)$ is the general solution of the homogeneous part \eqref{AngularPerturbations5} of \eqref{AngularPerturbations10} and is thus given by a formula of the form \eqref{GeneralSolution1}, whereas $\zeta_{k}^{\gamma}$ is a special solution. Plugging \eqref{GeneralSolution3} into \eqref{AngularPerturbations9} and using the fact that $\tilde{\eta}_{k}^{\gamma}\left(t\right)$ solves the homogeneous equation \eqref{AngularPerturbations4}, we get:
\begin{IEEEeqnarray}{l}
\left(\frac{2}{9} + \omega_3^2\right) \zeta_{i} - u_0\left(u_0 T_{ik} + Q_{ik}\right) \zeta_{k} = \tilde{f}_i, \qquad \label{Forcing2}
\end{IEEEeqnarray}
where we have set $F_k^{2\gamma} \equiv \tilde{f}_k^{\gamma} e^{\sqrt{2}t/3}$ in \eqref{Forcing1a}--\eqref{Forcing1c} so that the coefficients $\tilde{f}_k^{\gamma}$ can be directly read-off from the forcing. It is easy to work out the solution of \eqref{Forcing2},
\begin{IEEEeqnarray}{l}
\zeta_x^{3,\pm 1} = \pm \frac{9\xi^2}{5}\sqrt{\frac{3}{7\pi}}, \label{Solutionj3a} \\
\zeta_y^{3,\pm 1} = -\frac{9i\xi^2}{5}\sqrt{\frac{3}{7\pi}}, \label{Solutionj3b} \\
\zeta_z^{3,0} = -\frac{27\xi^2}{5\sqrt{7\pi}}. \label{Solutionj3c}
\end{IEEEeqnarray}
The NLO initial conditions \eqref{InitialConditionPerturbations2} imply
\begin{IEEEeqnarray}{l}
\eta_k^{2\gamma}\left(0\right) = \dot{\eta}_k^{2\gamma}\left(0\right) = 0,
\end{IEEEeqnarray}
which leads to the following set of constraints for the solution of the homogeneous equation $\tilde{\eta}_{k}^{\gamma}\left(t\right)$
\begin{IEEEeqnarray}{ll}
\tilde{\eta}_{x,y}^{3,\pm 1}\left(0\right) = -\zeta_{x,y}^{3,\pm 1}, \quad &\tilde{\eta}_z^{3,0}\left(0\right) = -\zeta_z^{3,0}, \\
\dot{\tilde{\eta}}_{x,y}^{3,\pm 1}\left(0\right) = -\frac{\sqrt{2}}{3}\zeta_{x,y}^{3,\pm 1}, \quad &\dot{\tilde{\eta}}_z^{3,0}\left(0\right) = -\frac{\sqrt{2}}{3}\zeta_z^{3,0}. \qquad
\end{IEEEeqnarray}
Based on the above, it can be shown that the solution \eqref{GeneralSolution3} satisfies the NLO Gauss law constraint \eqref{ConstraintSecondOrder} at the initial time $t=0$. \\
\indent The coupling of the $j = 2$ modes leads to a non-vanishing forcing $F_i^{2\gamma}$, not only in the case $j'' = 3$ (which we have just treated), but also when the spin becomes $j'' = 1$. Plugging \eqref{Solution1}--\eqref{Solution3} into the expression \eqref{ForcingOrder2} for the NLO forcing term, we obtain the following nonzero forcing components:
\begin{IEEEeqnarray}{l}
F_x^{2,1,1,\pm 1} = \pm\frac{3\xi^2}{40} \sqrt{\frac{3}{2\pi}}\cdot e^{\sqrt{2}t/3}, \label{Forcing1aj1} \\
F_y^{2,1,\pm 1} = -\frac{2i\xi^2}{40} \sqrt{\frac{3}{2\pi}}\cdot e^{\sqrt{2}t/3}, \label{Forcing1bj1} \\
F_z^{2,1,0} = -\frac{\xi^2}{10} \sqrt{\frac{3}{\pi}} \cdot e^{\sqrt{2}t/3}. \qquad \label{Forcing1cj1}
\end{IEEEeqnarray}
\noindent It is again straightforward to calculate the special solution of \eqref{Forcing2},
\begin{IEEEeqnarray}{l}
\zeta_x^{1,\pm 1} = \pm \frac{29\xi^2}{20\sqrt{6\pi}}, \label{Solutionj1a} \\
\zeta_y^{1,\pm 1} = -\frac{29i\xi^2}{20\sqrt{6\pi}}, \label{Solutionj1b} \\
\zeta_z^{1,0} = -\frac{8\xi^2}{5\sqrt{3\pi}}. \label{Solutionj1c}
\end{IEEEeqnarray}
\noindent The NLO initial conditions \eqref{InitialConditionPerturbations2} can again be enforced by choosing an appropriate solution $\tilde{\eta}_{k}^{\gamma}\left(t\right)$ of the homogeneous equation \eqref{AngularPerturbations5}. \\
\indent It is interesting to take a closer look at the decomposition of the above solutions into the three sectors $P$, $R_{\pm}$. As we have already mentioned, the LO mode $\xi^{2,0}_{-} \equiv \xi$ that we turned on belongs to the $R_-$ sector and corresponds to $j=2$ and $m=0$. It is a matter of algebra to show that the solutions \eqref{Solutionj3a}-\eqref{Solutionj3c} and \eqref{Solutionj1a}-\eqref{Solutionj1c} can be expressed in the $3\times(2j+1)$-dimensional space as follows:
\begin{equation}
Z_{3}=\frac{9\xi^2}{5}\sqrt{\frac{3}{\pi}}\cdot \vert - \rangle\Big|_{j=3, m=0},
\end{equation}
and
\begin{equation}
Z_{1}=-\frac{\xi^2}{10\sqrt{2\pi}}\cdot \vert + \rangle\Big|_{j=1, m=0} + \frac{3\xi^2}{2\sqrt{\pi}}\cdot \vert - \rangle\Big|_{j=1, m=0}, \qquad
\end{equation}
where we defined the $3\times(2j+1)$ vectors
\begin{IEEEeqnarray}{l}
Z_1 = \begin{pmatrix} \zeta_x^{1m} \\ \zeta_y^{1m} \\ \zeta_z^{1m} \end{pmatrix}, \qquad Z_3 = \begin{pmatrix} \zeta_x^{3m} \\ \zeta_y^{3m} \\ \zeta_z^{3m} \end{pmatrix}.
\end{IEEEeqnarray}
Therefore the instabilities do not only propagate to NLO modes of higher angular momenta $j$ (compared to the LO instabilities), but also between sectors. Note however that the magnitude of the $R_+$ instability is suppressed since $\vert\zeta^3_-\vert\simeq 44\vert\zeta^1_+\vert$ and $\vert\zeta^1_-\vert\simeq 21\vert\zeta^1_+\vert$.
\section[Conclusions and discussion]{Conclusions and discussion\label{Section:ConclusionsDiscussion}}
\noindent Motivated by the idea that the dynamics of the microscopic degrees of freedom on the horizon of static spherically symmetric black holes can be described by the BMN matrix model (a highly nonlocal field theory), we study the chaotic properties of this theory's classical continuum limit, that is super M2-brane theory in the background \eqref{MaximallySupersymmetricMetric}--\eqref{MaximallySupersymmetricMetricFieldStrength}. For previous work on this topic see \cite{AsanoKawaiYoshida15}. The solution \eqref{MaximallySupersymmetricMetric}--\eqref{MaximallySupersymmetricMetricFieldStrength} is nothing more that the Penrose-G\"{u}ven \cite{Penrose76, Gueven00} limit of the maximally supersymmetric backgrounds AdS$_{7,4}\times \text{S}^{4,7}$ of 11-dimensional supergravity. Because of the flux term \eqref{MaximallySupersymmetricMetricFieldStrength}, the corresponding supermembrane theory is found to contain both stable and unstable sectors. In the present paper we study the NLO perturbative dynamics of classical solutions of spherical topology in the $SO(3)$ sector of the continuum limit of the BMN matrix model. \\
\indent We show that the LO instabilities of certain spherical solutions (which appear exclusively in the dipole $j=1$ and quadrupole $j=2$ sectors of multipole perturbations, see \cite{AxenidesFloratosLinardopoulos17b}) give rise to a cascade of NLO instabilities that extend to all multipole sectors. The transmission of instabilities is due to the nonlinear coupling of NLO perturbations, which is in turn induced by the infinite-dimensional area-preserving symmetry of the membrane. The instability cascade is further enhanced at subsequent perturbative orders. The physical reason behind the development of instabilities is related to the infinite-dimensional symmetry of relativistic membranes. In other words, the development of infinitely many thin tubes only costs an infinitesimal amount of energy to the membrane. These thin tubes correspond to large values of the angular momentum quantum number $j$ arising in the expansion of perturbations in the multipole basis. This confirms our expectations that the unstable solutions ultimately induce a hedgehog structure on the membrane. In order to regularize these instabilities one has to rely on the matrix model truncation of the membrane (i.e.\ the BMN matrix model) where the dimension of the matrices $N$ determines a maximum value for each $j < N-1$. \\
\indent Concerning the chaotic properties of BMN membranes, the matrix model has a higher degree of nonlocality and is therefore expected to scramble perturbations faster than the continuous membrane (which contains a local Poisson bracket interaction term). Another exciting aspect relating to the development of membrane instabilities is the possibility of topology changes that are caused by the self-interactions of the membranes. This is a longstanding problem in the field and is closely related to our understanding of membrane field theory. Presently, only Euclidean-time solutions have been shown to be able to induce topology changes in membranes (see e.g.\ \cite{BerensteinDzienkowskiLashof-Regas15, KovacsSatoShimada15}). Because matrix theory is a field theory of discrete membranes (i.e.\ seen as bound states of D0-branes \cite{BFSS97}) it stands much better chances of providing an answer to this riddle. \\
\indent A straightforward extension of our present work is the study of NLO perturbations of the BMN membrane in the $SO(3)\times SO(6)$ sector. The $SO(6)$ rotating configuration encodes the effect of higher dimensions on the $3+1$-dimensional fluid system.
\section*{Acknowledgements}
\noindent The authors would like to thank C.\ Bachas, C.\ Efthymiopoulos, J.\ Hoppe, J.\ Iliopoulos and S.\ Nicolis for illuminating discussions. \\
\indent The present research is funded in the context of the project "Chaotic dynamics and black holes in BMN theory" E-12386 (MIS 5047794) under the call for proposals "Supporting researchers with an emphasis on young researchers---Cycle B" (EDULLL 103). The project is co-financed by Greece and the European Union (European Social Fund---ESF) by the Operational Programme Human Resources Development, Education and Lifelong Learning 2014-2020.
\appendix
\section[Structure constants]{Structure constants \label{Appendix:StructureConstants}}
\noindent The structure constants $f_{\alpha\beta}^{\gamma}$ of spherical harmonics that show up in the forcing term \eqref{ForcingOrder2} were defined in \eqref{SphericalHarmonics2}. It is rather straightforward to invert \eqref{SphericalHarmonics2},
\begin{equation}
f_{\alpha\beta}^{\gamma} = \int_{\text{S}^2} Y^{*}_{\gamma}\left(\theta,\phi\right)\left\{Y_{\alpha}\left(\theta,\phi\right),Y_{\beta}\left(\theta,\phi\right)\right\}d\Omega, \label{StructureConstants1}
\end{equation}
in order to obtain a closed formula for $f_{\alpha\beta}^{\gamma}$ that is valid for all the values of the quantum numbers $\alpha \equiv jm$, $\beta \equiv j'm'$ and $\gamma \equiv j''m''$. In the present paper, we only need the first few of them (see also \cite{ArakelianSavvidy89}):
\begin{widetext}
\begin{IEEEeqnarray}{l}
f^{j'm'}_{1,\pm 1;jm} = \pm i \, \sqrt{\frac{3}{8\pi}} \cdot \sqrt{\left(j\mp m\right) \left(j \pm m + 1\right)}\,\delta_{j'j}\,\delta_{m',m\pm 1}, \qquad
f^{j'm'}_{1,0;jm} = -i m \cdot\sqrt{\frac{3}{4\pi}} \, \delta_{j'j}\,\delta_{m'm} \qquad \label{StructureConstants2} \\[12pt]
f^{j'm'}_{2,0;jm} = -3i m \sqrt{\frac{5}{4\pi}} \cdot \left[\sqrt{\frac{\left(j + 1\right)^2 - m^2}{\left(2j + 1\right) \left(2j + 3\right)}} \, \delta_{j',j+1} + \sqrt{\frac{j^2 - m^2}{\left(2j + 1\right) \left(2j - 1\right)}} \, \delta_{j',j-1}\right]\cdot\delta_{m'm} \qquad \label{StructureConstants3} \\[12pt]
f^{j'm'}_{2,\pm 1;jm} = \pm i \sqrt{\frac{15}{8\pi}} \cdot \Bigg[\left(j \mp 2m\right)\cdot \sqrt{\frac{\left(j \pm m + 1\right) \left(j \pm m + 2\right)}{\left(2j + 1\right) \left(2j + 3\right)}} \, \delta_{j',j+1} + \nonumber \\
\hspace{6cm} + \left(j \pm 2m + 1\right) \cdot \sqrt{\frac{\left(j \mp m - 1\right) \left(j \mp m\right)}{\left(2j + 1\right) \left(2j - 1\right)}} \, \delta_{j',j-1}\Bigg] \cdot \delta_{m',m \pm 1} \qquad \label{StructureConstants4} \\[12pt]
f^{j'm'}_{2,\pm2;jm} = \pm i \sqrt{\frac{15}{8\pi}}\cdot \Bigg[\sqrt{\frac{\left(j \mp m\right) \left(j \pm m + 1\right) \left(j \pm m + 2\right) \left(j \pm m + 3\right)}{\left(2j + 1\right) \left(2j + 3\right)}}\,\delta_{j',j+1} - \nonumber \\
\hspace{5cm} - \sqrt{\frac{\left(j \mp m\right) \left(j \mp m - 1\right) \left(j \mp m - 2\right) \left(j \pm m + 1\right)}{\left(2j + 1\right) \left(2j - 1\right)}}\,\delta_{j',j-1}\Bigg]\cdot\delta_{m',m \pm 2}. \qquad \label{StructureConstants5}
\end{IEEEeqnarray}
\end{widetext}
\eqref{StructureConstants1} implies that the structure constants $f_{\alpha\beta}^{\gamma}$ obey the following sum rules:
\begin{equation}
m + m' = m'', \qquad j + j' + j'' = \text{odd},
\end{equation}
where the second equation is obtained by setting $(\theta,\phi)\rightarrow(\pi-\theta,\phi+\pi)$ in \eqref{StructureConstants1}. In addition $j,\thinspace j'$ and $j''$ obey triangle inequalities, e.g.\
\begin{equation}
\left|j - j'\right| + 1 \leq j'' \leq \left|j + j'\right| - 1,
\end{equation}
and its cyclic permutations hold. For the purposes of our work, this implies that when modes $n^{1jm}$ are turned on up to $j_{\textrm{max}}$, the forcing term vanishes for $F^{2j''m''}$ for $j''\geq 2j_{\textrm{max}}$.

\noindent
\bibliographystyle{JHEP}
\bibliography{HEP_Bibliography,Math_Bibliography}

\providecommand{\href}[2]{#2}\begingroup\raggedright\begin{thebibliography}{10}

\bibitem{SekinoSusskind08}
Y.~Sekino and L.~Susskind, {\it {Fast scramblers}},  {\em JHEP} {\bf
  \textbf{10}} (2008) 065, [\href{http://arxiv.org/abs/0808.2096}{{\tt
  arXiv:0808.2096}}].

\bibitem{Damour78}
T.~Damour, {\it {Black hole eddy currents}},  {\em Phys. Rev.} {\bf
  \textbf{D18}} (1978) 3598.

\bibitem{ThornePriceMacdonald86}
K.~S. Thorne, R.~H. Price, and D.~A. Macdonald, {\em {Black holes: The membrane
  paradigm}}.
\newblock 1986.

\bibitem{Hoppe82}
J.~Hoppe, {\em \href{http://dspace.mit.edu/handle/1721.1/15717}{Quantum theory
  of a massless relativistic surface and a two-dimensional bound state
  problem}}.
\newblock PhD thesis, Massachusetts Institute of Technology, 1982.

\bibitem{DasguptaJabbariRaamsdonk02}
K.~Dasgupta, M.~M. Sheikh-Jabbari, and M.~{Van Raamsdonk}, {\it {Matrix
  perturbation theory for M-theory on a pp-wave}},  {\em JHEP} {\bf
  \textbf{05}} (2002) 56, [\href{http://arxiv.org/abs/hep-th/0205185}{{\tt
  hep-th/0205185}}].

\bibitem{AxenidesFloratosLinardopoulos17a}
M.~Axenides, E.~Floratos, and G.~Linardopoulos, {\it {M2-brane dynamics in the
  classical limit of the BMN matrix model}},  {\em Phys. Lett.} {\bf
  \textbf{B773}} (2017) 265, [\href{http://arxiv.org/abs/1707.02878}{{\tt
  arXiv:1707.02878}}].

\bibitem{AxenidesFloratosLinardopoulos17b}
M.~Axenides, E.~Floratos, and G.~Linardopoulos, {\it {Multipole stability of
  spinning M2 branes in the classical limit of the BMN matrix model}},  {\em
  Phys. Rev.} {\bf \textbf{D97}} (2018) 126019,
  [\href{http://arxiv.org/abs/1712.06544}{{\tt arXiv:1712.06544}}].

\bibitem{AxenidesFloratosKatsinisLinardopoulos20}
M.~Axenides, E.~Floratos, D.~Katsinis, and G.~Linardopoulos, {\it {M-theory as
  a dynamical system generator}},  {\em Springer Proceedings in Complexity}
  (2020) in press, [\href{http://arxiv.org/abs/2007.07028}{{\tt
  arXiv:2007.07028}}].

\bibitem{Frisch95}
U.~Frisch, {\em {Turbulence: The legacy of A.\ N.\ Kolmogorov}}.
\newblock Cambridge University Press, 1995.

\bibitem{Penrose76}
R.~Penrose, {\it {Any space-time has a plane wave as a limit}},  in {\em
  {Differential geometry and relativity. A volume in honour of {Andr\'{e}}
  Lichnerowicz on his 60th birthday}} (M.~Cahen and M.~Flato, eds.), vol.~3 of
  {\em Mathematical Physics and Applied Mathematics}, p.~271.
\newblock Springer Netherlands, 1976.

\bibitem{OFarrillPapadopoulos02b}
J.~M. {Figueroa-O'Farrill} and G.~Papadopoulos, {\it {Maximally supersymmetric
  solutions of ten-dimensional and eleven-dimensional supergravities}},  {\em
  JHEP} {\bf \textbf{03}} (2003) 048,
  [\href{http://arxiv.org/abs/hep-th/0211089}{{\tt hep-th/0211089}}].

\bibitem{BMN02}
D.~Berenstein, J.~Maldacena, and H.~Nastase, {\it {Strings in flat space and pp
  waves from $\mathcal{N} = 4$ super Yang-Mills}},  {\em JHEP} {\bf
  \textbf{04}} (2002) 013, [\href{http://arxiv.org/abs/hep-th/0202021}{{\tt
  hep-th/0202021}}].

\bibitem{BFSS97}
T.~Banks, W.~Fischler, S.~H. Shenker, and L.~Susskind, {\it {M-theory as a
  matrix model: A conjecture}},  {\em Phys. Rev.} {\bf \textbf{D55}} (1997)
  5112, [\href{http://arxiv.org/abs/hep-th/9610043}{{\tt hep-th/9610043}}].

\bibitem{Myers99b}
R.~C. Myers, {\it {Dielectric branes}},  {\em JHEP} {\bf \textbf{12}} (1999)
  022, [\href{http://arxiv.org/abs/hep-th/9910053}{{\tt hep-th/9910053}}].

\bibitem{HarmarkSavvidy00}
T.~Harmark and K.~G. Savvidy, {\it {Ramond-Ramond field radiation from rotating
  ellipsoidal membranes}},  {\em Nucl. Phys.} {\bf \textbf{B585}} (2000) 567,
  [\href{http://arxiv.org/abs/hep-th/0002157}{{\tt hep-th/0002157}}].

\bibitem{CollinsTucker76}
P.~A. Collins and R.~W. Tucker, {\it {Classical and quantum mechanics of free
  relativistic membranes}},  {\em Nucl. Phys.} {\bf \textbf{B112}} (1976) 150.

\bibitem{EfstathiouSadovskii04}
K.~Efstathiou and D.~Sadovski\'{\i}, {\it {Perturbations of the 1:1:1 Resonance
  with tetrahedral symmetry: a three degree of freedom analogue of the two
  degree of freedom H\'{e}non–Heiles Hamiltonian}},  {\em Nonlinearity} {\bf
  \textbf{17}} (2004) 415.

\bibitem{Lefschetz57}
S.~Lefschetz, {\em {Differential equations: Geometric theory}}.
\newblock Wiley, 1957.

\bibitem{AxenidesFloratosPerivolaropoulos01}
M.~Axenides, E.~G. Floratos, and L.~Perivolaropoulos, {\it {Quadrupole
  instabilities of relativistic rotating membranes}},  {\em Phys. Rev.} {\bf
  \textbf{D64}} (2001) 107901, [\href{http://arxiv.org/abs/hep-th/0105292}{{\tt
  hep-th/0105292}}].

\bibitem{Rose57}
M.~E. Rose, {\em {Elementary theory of angular momentum}}.
\newblock Reidel, 1957.

\bibitem{AsanoKawaiYoshida15}
Y.~Asano, D.~Kawai, and K.~Yoshida, {\it {Chaos in the BMN matrix model}},
  {\em JHEP} {\bf \textbf{06}} (2015) 191,
  [\href{http://arxiv.org/abs/1503.04594}{{\tt arXiv:1503.04594}}].

\bibitem{Gueven00}
R.~{G\"{u}ven}, {\it {Plane-wave limits and T-duality}},  {\em Phys. Lett.}
  {\bf \textbf{B482}} (2000) 255,
  [\href{http://arxiv.org/abs/hep-th/0005061}{{\tt hep-th/0005061}}].

\bibitem{BerensteinDzienkowskiLashof-Regas15}
D.~Berenstein, E.~Dzienkowski, and R.~Lashof-Regas, {\it {Spinning the fuzzy
  sphere}},  {\em JHEP} {\bf \textbf{08}} (2015) 134,
  [\href{http://arxiv.org/abs/1506.01722}{{\tt arXiv:1506.01722}}].

\bibitem{KovacsSatoShimada15}
S.~Kovacs, Y.~Sato, and H.~Shimada, {\it {On membrane interactions and a
  three-dimensional analog of Riemann surfaces}},  {\em JHEP} {\bf \textbf{02}}
  (2016) 050, [\href{http://arxiv.org/abs/1508.03367}{{\tt arXiv:1508.03367}}].

\bibitem{ArakelianSavvidy89}
T.~A. Arakelian and G.~K. Savvidy, {\it {Geometry of a group of area-preserving
  diffeomorphisms}},  {\em Phys. Lett.} {\bf \textbf{B223}} (1989) 41.

\end{thebibliography}\endgroup

\end{document}